\documentclass[preprint,showpacs,preprintnumbers,amsmath,amssymb,epsf,
tightenlines]{revtex4}

\usepackage{graphicx}
\usepackage{subfigure}

\begin{document}

% Use the \preprint command to place your local institutional report
% number in the upper righthand corner of the title page in preprint mode.
% Multiple \preprint commands are allowed.
% Use the 'preprintnumbers' class option to override journal defaults
% to display numbers if necessary
\preprint{KAIST-TH 2002/11}

%Title of paper
\title{CP violations in the $K$ and $B$ meson systems \\
in the SUSY models with $(S_3)^3$ flavor symmetry}

\author{P. Ko}
\email[]{pko@muon.kaist.ac.kr}
%\homepage[]{}
%\thanks{}
%\altaffiliation{}
\affiliation{Dep. of Physics, KAIST, Daejon 305-701, Korea}

\author{Jae-hyeon\ Park}
\email[]{jhpark@muon.kaist.ac.kr}
\affiliation{Dep. of Physics, KAIST, Daejon 305-701, Korea}

\date{July 1, 2002}

\begin{abstract}
% insert abstract here
Rare decays of $B$ and $K$ mesons and CP violations thereof are considered 
in SUSY models with $S_3^3$ flavor symmetry.  SUSY contributions to
$\epsilon_K$ and $\epsilon^{'}/\epsilon_K$ can be large, but not to 
$B^0 - \overline{B^0}$ mixing because of strong constraint from $B\rightarrow
X_d \gamma$ for $( \delta_{13}^d )_{LR}$. Still large deviations from the
SM predictions are possible in the branching ratio for 
$B\rightarrow X_d \gamma$ and direct CP violations therein, even if the 
KM angle $\gamma$ is in the range preferred by the SM CKM fit. 
Experimental study of $B\rightarrow X_d \gamma$ is strongly recommended 
as a probe for new physics with nontrivial flavor structure in the $(13)$
mixing. 
\end{abstract}

% insert suggested PACS numbers in braces on next line
\pacs{}

\maketitle

%\section{Introduction}

%\subsection{}
%\subsubsection{}

By recent observations of $\epsilon^{'}/\epsilon_K$ and CP asymmetry in 
$B^0 (\overline{B^0})\rightarrow J/\psi K_S$, we now have a solid evidence 
that CP is indeed violated both in the $K$ and $B$ meson sectors, both in 
the mixing and in the decay. These observations are in accord with the SM 
predictions with a single CP violating KM phase. The current world averages 
of the relevant observables are summarized as follows \cite{pdg}:
\begin{eqnarray}
  \label{eq:data}
 ~~~ \epsilon_K~~~ & = & e^{i \pi/4}~( 2.280 \pm 0.013  ) \times 10^{-3}, 
\nonumber  %    \cite{pdg},
\\
  {\rm Re} ( \epsilon^{'} / \epsilon_K ) & = & 
( 19.2 \pm 2.4 ) \times 10^{-3},
%                    \cite{pdg},
\\
   \sin 2 \beta_{\psi K} & = & (0.79 \pm 0.10). 
\nonumber  %  \cite{pdg} 
\end{eqnarray}
The nonzero $\epsilon_K$ implies the CP violation in the $K^0-\overline{K^0}$ 
mixing, and provides a useful constraint on the CKM elements, albeit with some 
theoretical uncertainties related with nonperturbative QCD effects. 
The nonvanishing $\epsilon^{'} / \epsilon_K$ excludes the original form of
superweak model put forth by L. Wolfenstein \cite{wolfen}.  Also relatively
large hadronic uncertainties in the kaon physics forbid us from drawing
definite conclusions on the validity of the SM or necessity for new physics
to understand the observed $\epsilon^{'} / \epsilon_K$ \cite{review}.
The current world average of $\sin 2 \beta_{\psi K}$ indicates that the KM 
paradigm for CP violations in the SM is in a good shape, and it starts to 
provide a meaningful constraint on various new physics scenarios. 
For example, some SUSY models that were invented to explain the low value of 
$\sin 2 \beta_{\psi K}$ reported by BaBar at the  early stage are now being 
challenged by the updated data. 
We also include the dilepton charge asymmetry $A_{ll}$ and the 
$B_d \rightarrow X_d \gamma$ branching ratio constraint extracted from the 
recent experimental 
upper limit on the $B\rightarrow \rho\gamma$ branching ratio \cite{b2rho}
\[
B( B \rightarrow \rho \gamma ) < 2.3 \times 10^{-6}. 
\]
We will take $B ( B\rightarrow X_d \gamma ) < 1 \times 10^{-5}$ as a 
representative value.

In the modern language of effective field theory, it would be unnatural
to have a CP violation entirely from $| \Delta S | = 2$ four-fermion
operators. Any flavor changing ($\Delta F = 1,2$ with $F = S$ or $B$)
four-fermion operators can have CP violating phases in principle.
This has an important implication for the current construction of the  CKM
unitarity triangle, since two informations come from the $\Delta F = 2$ 
(with $F = S$ or $B$) processes, namely $\epsilon_K$ and $\Delta M_{B^0}$. 
In the presence of new physics that could give substantial contributions 
to the $\Delta F = 2$ mixing, there may be additional contributions to 
these quantities so that the current CKM global analysis could change.  
Of course, in the SM, three vastly different constraints from 
$b\rightarrow u l \nu$, $\epsilon_K$ and $\Delta M_{B^0}$ give a single 
overlapping region in the $( \rho, \eta )$ plane due to the heaviness of top 
quark. This is nontrivial at all, and could be regarded as a strong evidence 
that the CKM picture is really a correct description for the CP violations 
(at least within the flavor changing sector). However it is still premature 
to conclude that there are no new flavor or CP violations beyond those 
residing in the CKM matrix.

In particular, if we consider weak scale supersymmetric (SUSY) models as 
the new physics beyond the SM, one would expect generically large flavor and 
CP violations from the soft SUSY breaking terms. These SUSY flavor and SUSY 
CP problems put serious constraints on realistic model buildings. There are
different avenues for SUSY flavor problem : universality, decoupling and 
alignment using flavor symmetry. The universality assumption is usually taken
in the minimal supergravity model, and can be justified in the string models
with dilaton dominated SUSY breaking or in the gauge mediation scenarios and
some other scenarios. 
The phenomenology of this class of models are well studied. The general 
conclusions on the flavor physics are that the deviations from the SM 
predictions are rather small and will be hard to be observed considering 
experimental errors and theoretical uncertainties (mainly from QCD).
Another way to solve the SUSY flavor problem is the so called decoupling 
solution or effective SUSY models \cite{Cohen:1996vb}, 
in which the 1st/2nd generation squarks 
are very heavy ($> O(10)$ TeV) and almost degenerate, whereas third family 
squarks and gauginos are relatively light ($\lesssim 1$ TeV). The sleptons 
can be either heavy or light. In this case, the flavor sturctures of the 
sfermion masses and trilinear couplings are not much constrained, and there
is enough room for large deviations in the $B$ meson sector from gluino 
mediated $b\rightarrow s (d)$ transitions \cite{Cohen:1996sq}. 
However, it is fair to say that 
there is no well defined effective SUSY models in which various soft 
parameters appear with definite relations. This makes it hard to make definite
predictions in effective SUSY models, unlike the minimal SUGRA models or 
GMSB, etc.. Finally, the third way to solve SUSY flavor problem is to invoke 
the so--called alignment mechanism using some flavor symmetries which could 
be either abelian or nonabelian \cite{nir}. In this approach, flavor 
structures in soft terms and Yukawa couplings are treated on the equal 
footing from the outset. Therefore, one can understand both large hierarchies 
in the fermion masses and their mixings, and  suppressions of SUSY induced 
FCNC amplitudes. This may be aesthetically more attractive than assuming the 
universality of sfermion mass terms or decoupling solution, since the latter 
approach cannot shed light on the flavor problem in the Yukawa sectors. 
There are already many works in the literatures on the alignment, depending 
on the flavor symmetry groups that are chosen. Generic predictions within this
approach are larger effects on flavor physics without leading too much 
$K^0 - \overline{K^0}$, $B^0 - \overline{B^0}$ mixings and 
$\mu \rightarrow e \gamma$, etc. compared to the solution based
on the universality assumption.  Since there are no unique flavor models 
however, one has to rely on theoretical arguments and one's tastes to choose 
particular flavor groups and study their phenomenology.

In aligment mechanisms, there are still small residual misaligments between
the Yukawa couplings and sfermion mass matrices in the flavor space. 
When the Yukawa coupling is diagonalized, this misaligment leads to the 
flavor changing interactions involving a gluino.
The gluino mediated $b\rightarrow d$ transitions can be handled in the mass 
insertion approximation, in which the slight misaligment between fermion 
and sfermion  mass matrices is characterized by parameters 
$(\delta_{ij}^d )_{AB}$ where $i,j=1,2,3$ are flavor indices and
$A,B = L,R$ represent the chirality of their superpartner fermions.
In order to study any nontrivial flavor physics from SUSY sector, it is
essential to estimate typical sizes of $(\delta_{ij}^d )_{AB}$ with
$AB = LL, LR, RR$ and $RL$. The answer will depend on how to solve the
SUSY flavor/CP problems.

In this work, we consider a flavor group $S_3^Q \times S_3^U \times S_3^D
\equiv S_3^3$, which was considered by Hall and Murayama \cite{hm} sometime 
ago. In SUSY models with the $S_3^3$ flavor group, flavor structures of the 
Yukawa couplings in the superpotential and sfermion mass matrices in the soft
SUSY breaking terms are both controlled by the same flavor symmetry group. 
The $S_3$ group acts on three elements $(1,2,3)$, and has six 
elements, $S_3 = { e, (12), (13), (23), (123), (132) }$, with $e$ being 
the identity element. The group $S_3$ has the following irreducible 
representations : {\bf 1}$_A$, {\bf 2} and {\bf 1}$_S$.  {\bf 1}$_A$ keeps 
(flips) its sign under even (odd) permutations of $(1,2,3)$, and {\bf 1}$_S$ 
is a trivial representation of $e$. It is assumed 
that this flavor symmetry is a local symmetry, in order that quantum gravity 
effects do not spoil it. Furthermore, the flavor symmetry is taken to be a 
local discrete symmetry. Otherwise, there would appear dangerous family 
dependent $D$ term contributions to sfermion masses after gauged flavor 
symmetry is broken to generate Yukawa couplings. 

In this model, the 1st/2nd generations are taken to transform as the
{\bf 2}, whereas the 3rd generation transforms as {\bf 1}$_A$ in order to 
make discrete flavor guage symmetry anomaly free. Finally one has to 
introduce three independent $S_3$ groups for $Q, U, D$ superfields in the 
MSSM in order to have a top quark the only massive particle in the flavor 
symmetry limit.
Both Higgs doublets $H_u, H_d$ transform as ( {\bf 1}$_A$, {\bf 1}$_A$,
{\bf 1}$_S$ ) under $S_3^Q \times S_3^U \times S_3^D$. Based on these 
assignments, one can construct the Yukawa and sfermion mass matrices. 
After diagonalizing the Yukawa matrices, one finds that typical sizes of 
mass insertion parameters $(\delta_{ij}^d )_{AB}$ are given by 
(see also Table~\ref{tab1}) 
\begin{equation}
( \delta_{12}^d )_{LL},~ %\sim %& O ( \lambda^3 ) \sim O (10^{-2} - 10^{-3}),
%\nonumber   \\
( \delta_{13}^d )_{LL} %\sim %& O ( \lambda^3 ),
%\nonumber   \\
% ( \delta_{23}^d )_{LL} 
\sim  O ( \lambda^3 ) \sim O (10^{-2} - 10^{-3}),
\end{equation}
whereas the typical sizes of $(\delta_{ij}^d )_{LR}$ insertions are
\begin{equation}
( \delta_{12} )_{LR} \sim \lambda^5,~~~~
%\nonumber   \\
( \delta_{13}^d )_{LR} %& \sim & O ( \lambda^3 ),
%\nonumber   \\
% ( \delta_{23}^d )_{LR}  
\sim  O ( \lambda^3 ).
\end{equation}
\begin{table}%[H] add [H] placement to break table across pages
\caption{\label{tab1}Typical sizes of $( \delta_{AB}^d )_{ij}$'s
in the $( S_3 )^3$
model. The row and the column denote the chiralities $AB$ and the family 
indices $ij$, respectively. $h$'s are Yukawa couplings, 
$m_t = h_t \langle H_u \rangle$ and $m_b = h_b \langle H_d \rangle$, etc.,
and $\lambda = 0.22$ is the sine of Cabbibo angle. 
The numbers in the parentheses are the limits on the quantities 
(for $\tilde{m} = m_{\tilde{g}} = 500$ GeV) derived from various low 
energy data such as neutral meson mixings,
$\epsilon^{'} / \epsilon_K$ and $b\rightarrow s \gamma$, etc.~ 
\cite{gabriel}. }
\begin{ruledtabular}
\begin{tabular}{ccc}
 & $(12)$ & $(13)$ %& $(23)$ 
\\
\hline 
%\\
$(LL)$ & $ h_t A \lambda^3 $ & $ h_t A \lambda^3 $ %& $ A \lambda^3 (??)$
\\
       & (0.05)              & (0.1)  %&  () 
\\
$(RR)$ & $ h_s^2 \lambda $ & $h_s^2 \lambda$  %& $ (??)  $
\\
       & (0.05)              & (0.1)  %&  ()
\\
$(LR)$ & $ h_s^2 \lambda $ & $ h_b h_t A \lambda^3 $ %& $ (??) $
\\
       & (0.008)             & (0.06)  %& ()
\end{tabular}
\end{ruledtabular}
\end{table}
Other parameters are typically much smaller $\lesssim O( \lambda^4 )$. 
Then, $( \delta^d_{12} )_{LL} \sim O(10^{-2} - 10^{-3})$ can generate both
$\epsilon_K$, ${\rm Re} ( \epsilon^{'} /\epsilon_K )$  as discussed by us
\cite{ko,ko2}. On the other hand, for $B^0 - \overline{B^0}$ mixing, 
both $( \delta_{13}^d )_{LL}$ and $( \delta_{13}^d )_{LR}$ should be 
considered together, since they are of the same sizes of $O ( \lambda^3 )$
for large $\tan \beta$
\footnote{Another motivation for choosing the $S_3^3$ flavor group was the 
claim that $( \delta_{13}^d )_{LL} \sim O(\lambda^2)$,
which could lead to large
SUSY CP violations in the $B$ meson sector. If their claim were true, SUSY 
contribution from $ ( \delta_{13}^d )_{LL}$ could saturate the 
$B^0 - \overline{B^0}$ mixing and CP asymmetry in 
$B^0 \rightarrow J/\psi K_s$. Then CP violations in both $K$ and $B$ meson 
sectors could be dominated by SUSY contributions. Unfortunately, we could 
not confirm their claim. Instead, we found 
$( \delta_{13}^d )_{LL}  \sim  O ( \lambda^3 )$, which is too small to 
saturate CP violations in the $B$ meson sector. Also it has a similar size 
as the $LR$ insertion, which is strongly constrained by 
$B\rightarrow X_d \gamma$ as discussed in the following.}.
Each element could have a phase of $\sim O(1)$, leading to new CP violations 
on top of the KM phase. Therefore, it is important to keep both 
$( \delta_{13}^d )_{LL}$ and $( \delta_{13}^d )_{LR}$, when we consider
the $B^0 - \overline{B^0}$ mixing, $\sin 2\beta_{\psi K}$, the dilepton
CP asymmetry $A_{ll}$ and $B\rightarrow X_d \gamma$. 
Since $B^0 - \overline{B^0}$ mixing can be saturated by gluino  mediated
amplitude only if $( \delta_{13}^d )_{LL} \sim O(10^{-1}) \sim O(\lambda)$ 
or $( \delta_{13}^d )_{LR} \sim 10^{-2}$, it is not possible to have 
significant contributions to $B^0 - \overline{B^0}$ mixing from 
$( \delta_{13}^d )_{LL} \sim O(\lambda^3)$ only in $S_3^3$ model. The $LR$ 
insertion parameter $( \delta_{13}^d )_{LR} \sim 10^{-2}$ is at the right 
range to saturate $B^0 - \overline{B^0}$ mixing in the $S_3^3$ model. 
But it turns out that such a large $( \delta_{13}^d )_{LR}$ may 
overproduce $B_d \rightarrow X_d \gamma$, thus being strongly constrained 
\footnote{This point was overlooked in the recent analysis 
by Be\'{c}irevi\'{c} et al.~\cite{rome}
(see Ref.~\cite{ko3} for detailed model independent analysis.}.
In short, there could be large SUSY contributions to the $\epsilon_K$ and
$\epsilon^{'} / \epsilon_K$, but not in the $B^0 - \overline{B^0}$ mixing,
if we consider $B\rightarrow X_d \gamma$ constraint. Still we find that 
there could be large deviations in the $B_d \rightarrow X_d \gamma$
branching ratio and the direct CP violation therein compared to the SM 
predictions. Thus the detailed experimental study of 
$B_d \rightarrow X_d \gamma$ can provide us with the flavor structure of 
the sfermion mass matrices, thereby a hint for a possible solution to 
the SUSY flavor problem. 

With these comments in mind, let us first consider the CP violation in the 
kaon sector. It is well known that generic SUSY models will have too large
FCNC amplitudes, unless  $(\delta_{ij}^d )_{AB}$'s are small enough
\cite{gabriel,masiero}. This is nothing but the SUSY flavor/$\epsilon_K$
problem \cite{nir}. A simple twist of this  observation is that the SUSY
contribution can be saturated by $(\delta_{ij}^d )_{AB}$.  In particular,
the observed $\epsilon_K$ can be saturated by
$(\delta_{12}^d )_{LL} \sim O(10^{-3})$. Masiero and Murayama showed that 
$(\delta_{12}^d )_{LR} \sim O(10^{-5})$ is possible in general SUSY models, 
and can explain the observed $\epsilon^{'}/\epsilon_K$ \cite{mm}. 
Also, flavor nonuniversal trilinear couplings can lead to large 
$\epsilon^{'}/\epsilon_K$ \cite{khalil}. In these cases, $\epsilon_K$
and $\epsilon^{'}/\epsilon_K$ are generated by two different parameters.
On the other hand, the present authors showed that the parameter
$(\delta_{12}^d)_{LL}$ can also  saturate and $\epsilon^{'}/\epsilon_K$ in
the double mass insertion approximation, if $| \mu \tan\beta | \sim O(10)$
TeV \cite{ko}. Thus a single SUSY parameter $(\delta_{12}^d)_{LL} \sim 10^{-2}
- 10^{-3}$ with a phase $\sim O(1)$ can generate (dominant portions of) 
both $\epsilon_K$ and $\epsilon^{'}/\epsilon_K$. Since there could be large
SUSY contributions in the $S_3^3$ model, the usual CKM phenomenology should 
be altered and the KM angle $\gamma$ would not be strongly constrained
by $\epsilon_K$ any longer. The implications of this scenario for other rare 
kaon decays such as $K\rightarrow \pi \nu \bar{\nu}$ and $K_L \rightarrow
\pi^0 e^+ e^-$ were studied in detail in Ref.~\cite{ko2}. 

The most general effective Hamiltonian for $B^0 -\overline{B^0}$ mixing
($\Delta B = 2$) can be written as the following form :
\begin{equation}
H_{\rm eff}^{\Delta B = 2} = \sum_{i=1}^5 C_i Q_i + 
\sum_{i=1}^3 \tilde{C}_i \tilde{Q}_i, 
\end{equation}
where the operators $Q_i$'s are defined as
\begin{eqnarray}
Q_1 & = &  \bar{d}_L^{\alpha} \gamma_\mu b_L^{\alpha}~
           \bar{d}_L^{\beta}  \gamma^\mu b_L^{\beta}\,
\nonumber  \\
\tilde{Q}_2 & = &  
\bar{d}_L^{\alpha} b_R^{\alpha}~\bar{d}_L^{\beta} b_R^{\beta}\, 
\nonumber  \\
\tilde{Q}_3 & = &  
\bar{d}_L^{\alpha} b_R^{\beta} ~\bar{d}_L^{\beta} b_R^{\alpha}\, 
%\nonumber  \\
%Q_4 & = &  \bar{d}_R^{\alpha} b_L^{\alpha}~\bar{d}_L^{\beta} b_R^{\beta}\, 
%\nonumber  \\
%Q_5 & = &  \bar{d}_R^{\alpha} b_L^{\beta} ~\bar{d}_L^{\beta} b_R^{\alpha}\,
\end{eqnarray}
$\alpha, \beta$ are color indices, and $q_{L,R} \equiv (1\mp \gamma_5) q /2$. 

The Wilson coefficients $C_i$'s in Eq.~(4) are obtained by calculating 
the $t-W$ in the SM and $\tilde{g}-\tilde{q}$ box diagrams in the mass 
insertion approximations in general SUSY models :
\begin{eqnarray}
C_1 & = & - {\alpha_s^2 \over 216 \tilde{m}^2}~\left( 24 x f_6 (x) + 66 
\tilde{f}_6 (x)\right) ~\left( \delta_{13}^d \right)_{LL}^2 \,
\nonumber  \\
\tilde{C}_2 & = & - {\alpha_s^2 \over 216 \tilde{m}^2}~204 x f_6 (x) 
~\left( \delta_{13}^d \right)_{LR}^2 \,
\nonumber  \\
\tilde{C}_3 & = & {\alpha_s^2 \over 216 \tilde{m}^2}~36 x f_6 (x) 
~\left( \delta_{13}^d \right)_{LR}^2 \,
\end{eqnarray}
Other Wilson coefficients are zero in the $S_3^3$ models ignoring 
$( \delta_{13}^d )_{RR}$ and $( \delta_{13}^d )_{RL}$. 
The SM contribution generate only operator $Q_1$, and the corresponding 
Wilson coefficient $C_1^{\rm SM}$ is given by \cite{Buras:1998ra}
\begin{equation}
C_1^{\rm SM} = 
\frac{G_F^2}{4 \pi^2} M_W^2 (V_{td}^* V_{tb})^2 S_0(x_t) ,
\end{equation}
where
\begin{equation}
  S_0(x_t) = \frac{4 x_t - 11 x_t^2 + x_t^3}{4 (1 - x_t)^2} 
  - \frac{3 x_t^3 \ln x_t}{2 (1 - x_t)^3} ,
\end{equation}
with $x_t \equiv m_t^2 / m_W^2$.
These Wilson coefficients for SUSY and SM are calculated at 
$\mu \sim m_{\tilde{g}} \sim \tilde{m}$ and $m_t$, respectively, and 
the RG running between these two scales will be ignored. 
Here $\tilde{m}$ is the common squark mass used in the mass insertion 
approximation, and $x \equiv m_{\tilde{g}}^2 / \tilde{m}^2$. 
The loop functions $f_6 (x)$ and $\tilde{f}_6 (x)$ are given by
\begin{eqnarray}
f_6 (x) & = & {6 ( 1 + 3 x ) \ln x + x^3 - 9 x^2 - 9 x + 17 \over 
               6 ( x - 1 )^5 }\,
\nonumber \\
\tilde{f}_6 (x) & = & {6 x ( 1+x)\ln x - x^3 - 9 x^2 + 9 x + 1 \over
               3 ( x - 1 )^5 } .
\end{eqnarray}

The above $\Delta B=2$ effective Hamiltonian will contribute to $\Delta m_B$,
the dilepton CP asymmetry and time dependent CP asymmetry in the decay
$B\rightarrow J/\psi K_s$ via the phase of the $B^0 - \overline{B^0}$ mixing.
Defining the mixing matrix element by
\begin{equation}
M_{12} (B^0) \equiv {1\over 2 m_B}~\langle B^0 | H_{\rm eff}^{\Delta B = 2}
| \overline{B^0} \rangle \,  
\end{equation}
one has $\Delta m_{B_d} = 2 | M_{12} (B_d^0) |$ since this quantity is 
dominated by the short distance contributions, unlike the $\Delta m_K$ for 
which long distance contributions would be significant. Therefore the data 
on $\Delta m_{B_d}^{\rm exp}$ will constrain the modulus of $M_{12} (B_d^0)$.

When we take matrix elements for four quark operators between $B^0$ and 
$\overline{B^0}$ states, we use the lattice improved calculations for the bag
parameters: 
%[I'M NOT SURE IF THIS IS ABSOLUTELY NESSECARI ! WE MAY USE THE OLD
%VACUUM INSERTION APPROXIMATIONS ! BUT I INVITE YOUR SUGGESTIONS ] : 
\begin{eqnarray}
\langle B_d | Q_1 (\mu) | \overline{B^0} \rangle & = &
{2\over 3}~m_{B_d}^2 f_{B_d}^2 B_1 ( \mu ),
\nonumber  \\
\langle B_d | \tilde{Q}_2 (\mu) | \overline{B^0} \rangle & = &
-{5\over 12}~\left( { m_{B_d} \over m_b (\mu) + m_d (\mu) } \right)^2~
m_{B_d}^2 f_{B_d}^2 B_2 ( \mu ),
\nonumber  \\
\langle B_d | \tilde{Q}_3 (\mu) | \overline{B^0} \rangle & = &
{1\over 12}~\left( { m_{B_d} \over m_b (\mu) + m_d (\mu) } \right)^2~
m_{B_d}^2 f_{B_d}^2 B_3 ( \mu ),
%\nonumber  
%\\
%\langle B_d | Q_4 (\mu) | \overline{B^0} \rangle & = &
%{1\over 2}~\left( { m_{B_d} \over m_b (\mu) + m_d (\mu) } \right)^2~
%m_{B_d}^2 f_{B_d}^2 B_1 ( \mu ),
%\nonumber  \\
%\langle B_d | Q_5 (\mu) | \overline{B^0} \rangle & = &
%{1\over 6}~\left( { m_{B_d} \over m_b (\mu) + m_d (\mu) } \right)^2~
%5m_{B_d}^2 f_{B_d}^2 B_1 ( \mu ),
%\nonumber
\end{eqnarray}
where the $B$ parameters $B_i (\mu)$'s are given by \cite{Becirevic:2001xt}
\begin{equation}
B_1 ( m_b ) = 0.87(4)^{+5}_{-4},~~~ B_2 ( m_b ) = 0.82(3)(4),
%\nonumber \\
~~~B_3 ( m_b ) = 1.02(6)(9),  
%      && B_4 ( m_b ) = 1.16(3)^{+5}_{-7},
%\\
%B_5 ( m_b ) = 1.91(4)^{+22}_{-7}.
\end{equation}
Also, we use the following running quark masses in the RI-MOM scheme :
\begin{equation}
m_b ( m_b ) = 4.6 ~{\rm GeV},~~~~ m_d ( m_b ) = 5.4~{\rm GeV}.
\end{equation} 
The bottom quark mass is obtained from the $\overline{MS}$ mass 
$m_b^{\overline{MS}} ( m_b^{\overline{MS}} ) = 4.23$ GeV. 
For the $B_d$ meson decay constant, we assume $f_{B_d} = 200 \pm 30$ MeV.
We set the common squark mass to $\tilde{m} = 500$ GeV and $x=1$,  
and $|V_{cb}| = (40.7 \pm 1.9) \times 10^{-3}$,
$| V_{ub} | = (3.61 \pm 0.46 )\times 10^{-3}$ with $2 \sigma$ variation.  
There are also contributions from $( \delta_{13}^d )_{LR}^{\rm ind}$, 
which is the induced $LR$ mixing due to the double mass insertion :
\begin{equation}
  ( \delta_{13}^d )_{LR}^{\rm ind} = ( \delta_{13}^d )_{LL} \times
{m_b ( A_b - \mu \tan\beta ) \over \tilde{m}^2 },
\end{equation}
which could be important for large $| \mu\tan\beta | \sim O(5-10)$ TeV 
\cite{ko,ko2}. We numerically checked that this contribution of 
$( \delta_{13}^d )_{LR}^{\rm ind}$ to $M_{12}$ is smaller than the SUSY 
contribution we showed above, and we do not show the explicit form here.  
However its effect can be important for the radiative decay 
$B\rightarrow X_d \gamma$, for which we will keep the induced $LR$ mixing.

Since $\Delta M_B = (0.472 \pm 0.017) ~{\rm ps}^{-1}$ \cite{pdg} 
is dominated by the short distance physics, it can be reliably calculated in 
the perturbation theory and is equal to $2 | M_{12}^{\rm full} |$. 
The argument of $M_{12}^{\rm full}$ (denoted by $2 \beta_{\psi K}$) is 
determined by the time dependent CP asymmetry in 
$B^0 (\overline{B^0}) \rightarrow J/\psi K_S$ as long as this decay is 
dominated by the SM tree amplitude, which is still a good assumption in our 
model. In the SM, one has $2 \beta_{\psi K} = 2 \beta$.  
Since $M_{12}^{\rm SUSY}$ can carry additional phases due to complex 
parameters $( \delta_{13}^d )_{LL}$ and $( \delta_{13}^d )_{LR}$, these will 
interfere with  $M_{12}^{\rm SM}$, and its effect will appear in the net
$B^0 - \overline{B^0}$ mixing. If we fix the KM angle $\gamma$ to a certain 
value, the $\Delta M_B$ will give a relation between the modulus and the
phases of $( \delta_{13}^d )_{LL}$ and $( \delta_{13}^d )_{LR}$.  We varied 
their phases from 0 to $2\pi$, and their moduli from 0 to 0.06, since their 
sizes have the order of magnitude $\sim \lambda^3$ with a fuzzy factor of 
$\sim 1/5 - 5$ in the SUSY models with  $S_3^3$ flavor symmetry.
The procedure of parameter space searching is as follows. 
For a particular set of moduli of 
$( \delta_{13}^d )_{LL}$ and $( \delta_{13}^d )_{LR}$,
their phases are varied between 0 and $2 \pi$, 
$\sqrt{\rho^2 + \eta^2}$ within $2 \sigma$ range,
and $f_{B_d}$ between $200 \pm 30$ MeV.
If there exists a parameter set that gives 
$\Delta M_B$ and $\sin 2 \beta_{\psi K}$ within $2 \sigma$
from the central values of the measurements,
this set of moduli is considered to be consistent with the experiments,
and those parameters are used to compute other observables such as
$B(B\rightarrow X_d \gamma)$, $A_{ll}$, and
$A_{\rm CP}^{b\rightarrow d\gamma}$ discussed below.
Throughout the presentation,
a parameter set that gives 
% $B(B\rightarrow X_d \, g) > 6.8 \%$
$B ( B \rightarrow X_d \gamma ) > 1 \times 10^{-5}$ 
is marked by a light gray (magenta) point,
% $B(B\rightarrow X_d \, g) < 6.8 \%$ but 
% $B ( B \rightarrow X_d \gamma ) > 0.5 \times 10^{-4}$
% by dark gray (green),
and $B ( B \rightarrow X_d \gamma ) < 1 \times 10^{-5}$
by dark gray (blue).
% \footnote{In our case,
% no such point was found that
% $B(B\rightarrow X_d \, g) > 6.8 \%$ and
% $B ( B \rightarrow X_d \gamma ) < 0.5 \times 10^{-4}$.
% }

\begin{figure}
\begin{center}%
% \parbox{3in}{\epsfxsize=7.5cm \epsfbox{newfig/gam000-0TeV.eps}\\
% \centerline{\footnotesize(a) \(\gamma = 0^\circ\)}}%
% \hspace{.2in}%
% \parbox{3in}{\epsfxsize=7.5cm \epsfbox{newfig/gam060-0TeV.eps}\\
% \centerline{\footnotesize(b) \(\gamma = 60^\circ\)}}\\%
% \vspace{.2in}%
%\parbox{3in}{\epsfxsize=3in\epsfbox{sin2bp-2phi-0.eps}\\
%\centerline{\footnotesize(c) \(\gamma = 0^\circ\)}}%
%\hspace{.2in}%
%\parbox{3in}{\epsfxsize=3in\epsfbox{sin2bp-2phi-55.eps}\\
%\centerline{\footnotesize(d) \(\gamma = 55^\circ\)}}
  \subfigure[$\gamma = 0^\circ$]{%
    \includegraphics[width=7cm]{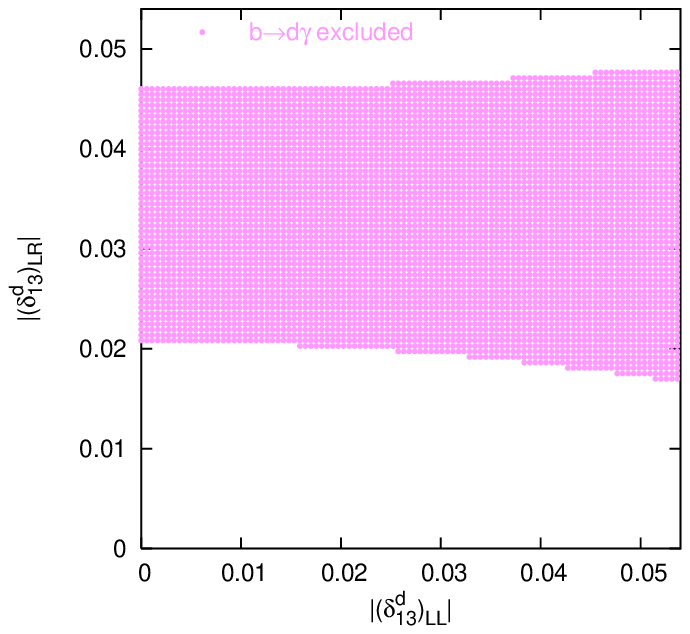}\hspace*{5mm}}
  \subfigure[$\gamma = 50^\circ$]{%
    \includegraphics[width=7cm]{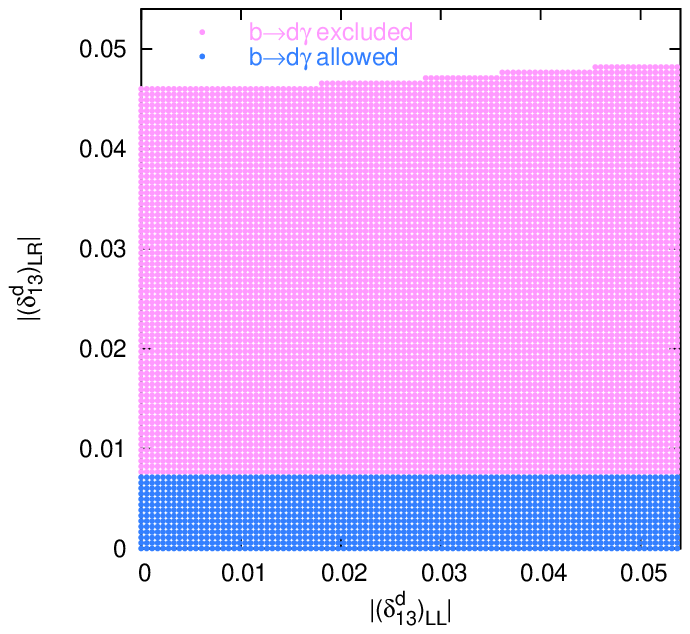}\hspace*{5mm}}
\end{center}
\caption{
The allowed region in the $( | ( \delta_{13}^d )_{LL} |,  
| ( \delta_{13}^d )_{LR} | )$ plane, which is consistent with $\Delta M_B$
and $a_{\psi K}$ measurements, for (a) $\gamma = 0$ and 
(b) $\gamma = 50^{\circ}$, respectively. 
Those parameters which lead to
% $B ( B \rightarrow X_d \, g ) > 6.8 \%$ 
$B ( B \rightarrow X_d \gamma ) > 1 \times 10^{-5}$ 
are represented by
light gray (magenta), 
% $B ( B \rightarrow X_d \, g) < 6.8 \%$ by
% dark gray (green),
and $B ( B \rightarrow X_d \gamma ) < 1 \times 10^{-5}$ by
dark gray (blue).
}
\label{fig1}
\end{figure}

In Figs.~\ref{fig1} (a) and (b), we show the allowed region in the 
$(| ( \delta_{13}^d )_{LL} |, | ( \delta_{13}^d )_{LR} | )$ plane, which
is consistent with the measured values of $\Delta M_B$ and 
$\sin 2\beta_{\psi K}$, for (a) $\gamma = 0^\circ$ and 
(b) $\gamma = 50^{\circ}$, respectively. 
% In both cases, the sizes of the 
% allowed $| ( \delta_{13}^d )_{LL} |$ and $| ( \delta_{13}^d )_{LR} |$ are 
% about $O( \lambda^3 )$, as predicted in the $S_3^3$ flavor symmetry. 
In both cases, there are allowed regions of
$| ( \delta_{13}^d )_{LL} |$ and $| ( \delta_{13}^d )_{LR} |$
with their magnitudes about $O( \lambda^3 )$, 
as predicted in the $S_3^3$ flavor symmetry. 
We also note that the observed $\sin 2 \beta_{\psi K} $ can be explained 
entirely by the SUSY effect (namely for $\gamma = 0^\circ$) for 
$(\delta_{13}^d)_{LR} \sim 10^{-2}$ which can be accommodated in our model. 
In other words, the real CKM is still consistent with the observed large CP 
asymmetry $a_{\psi K}$ in our model.  [Note that the SM prediction for this 
asymmetry is $\sin 2 \beta_{\psi K} = ( 0.698 \pm 0.066 )$.]  For this to be 
ture, it is crucial to have both $LL$ and $LR$ mixings in the $(13)$ sector 
of similar sizes $\sim \lambda^3$. If we neglected $LR$ mixing, then the size 
of the $(LL)$ mixing should be $\sim O( \lambda^2 - 10^{-1} )$ in order to 
saturate the measured $\Delta M_B$ and $a_{\psi K}$, which is too large to be 
accommodated within the $S_3^3$ flavor symmetry group. 
However, this is not the whole story, since the $( \delta_{13}^d )_{LR}$ 
parameter is strongly constrained by $B\rightarrow X_d \gamma$, just as 
the $( \delta_{23}^d )_{LR}$ parameter is strongly constrained by 
$B\rightarrow X_s \gamma$. Relegating the discussion on this radiative decay
to the later part of this work, we simply represent the parameter space with 
$B(B\rightarrow X_d \gamma) < (>) ~ 1 \times 10^{-5}$ by the dark (light) 
area. Then $\gamma=0^\circ$ is no longer possible. Namely fully SUSY CP
violations in the $B$ meson sector is not possible in the $S_3^3$ model,
unlike the $K$ meson sector. For $\gamma = 55^\circ$, there exists some 
parameter space where SUSY contributions to $B^0 - \overline{B^0}$ mixing 
is consistent with the current data and also with the SM case. 

\begin{figure}
% \centerline{\epsfxsize=10cm \epsfbox{}}
  \begin{center}
    \includegraphics[width=7cm]{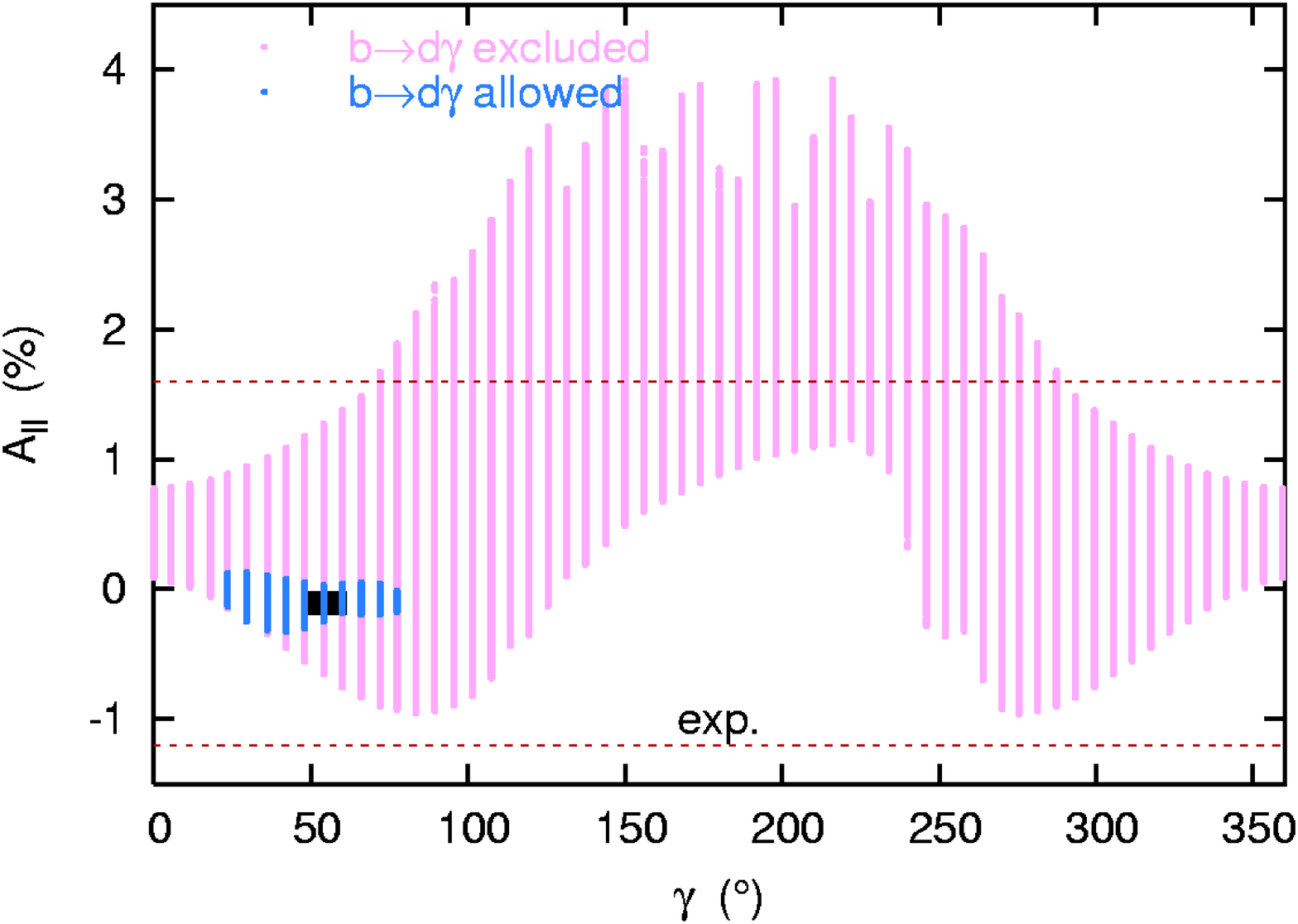}
  \end{center}
\caption{
  %The correlation between $a_{\psi K_S}$ and 
  $A_{ll}$ in in our model as a function of %for different values of 
  $\gamma$. 
  The vertical lines are the data.
  The black rectangle around $\gamma \simeq 55^\circ$ is
  the SM prediction.
% the SM prediction are represented by the vertical 
% line and the filled square, respectively. 
% Those parameters which lead to
% $B ( B_d \rightarrow X_d \gamma ) > 0.5 \times 10^{-4}$ are represented by
% the crosses.
Those parameters which lead to
% $B ( B \rightarrow X_d \, g ) > 6.8 \%$ 
$B ( B \rightarrow X_d \gamma ) > 1 \times 10^{-5}$ 
are represented by
light gray (magenta), 
% $B ( B \rightarrow X_d \, g) < 6.8 \%$ by
% dark gray (green),
and $B ( B \rightarrow X_d \gamma ) < 1 \times 10^{-5}$ by
dark gray (blue).
}
\label{fig2}
\end{figure}

The new parameters $( \delta_{13}^d )_{LL, LR}$ would affect the dilepton 
asymmetry $A_{ll}$ through $B^0 - \overline{B^0}$ mixing 
\cite{Randall:1998te}:
%, which is proportional to ${\rm Re} ( \epsilon_B )$ :
\[
A_{ll} \equiv {N(BB) - N(\bar{B}\bar{B}) \over N(BB) + N(\bar{B}\bar{B})}
%= - {| \epsilon_B |^4 - 1 \over | \epsilon_B |^4 + 1 }
%= {{\rm Im} (\Gamma_{12} / M_{12} ) \over 1 + | \Gamma_{12} / M_{12} |^2 / 4 }
\approx {\rm Im} (\Gamma_{12} / M_{12} ).
\]
Here $M_{12}, \Gamma_{12}$ are the matrix element of $B-\overline{B^0}$ mixing
\[
\langle \overline{B} | H | B \rangle = M_{12} - {i \over 2} \Gamma_{12}.
\]
In the SM, the phases of $M_{12}$ and $\Gamma_{12}$ are approximately equal
and
\[
\Delta M_{\rm SM} \approx 2 | M_{12}^{\rm SM} |, ~~~~
\Delta \Gamma_{\rm SM} \approx 2 | \Gamma_{12}^{\rm SM} |.
\]
% Using the most recent determination of $(\rho, \eta)$, we get
Using $42.4^\circ \leq \gamma \leq 67.2^\circ$ \cite{Ciuchini:2000de}
and the other parameters in the same range as is used in the figures,
we get
\[
% -1.28 \times 10^{-3} \leq A_{ll}^{\rm SM} \leq -0.48 \times 10^{-3},
-1.98 \times 10^{-3} \leq A_{ll}^{\rm SM} \leq -0.16 \times 10^{-3},
\] 
whereas the current world average is 
\[
A_{ll}^{\rm exp} \approx (0.2 \pm 1.4) \times 10^{-2}.
\]
In the presence of SUSY, the phases of $M_{12}$ and $\Gamma_{12}$ may be no
longer the same, and potentially larger dilepton asymmetry may be possible.
In particular, $M_{12}$ could be affected a lot by SUSY particles, whereas
$\Gamma_{12}$ is not (since it would be at higher order) :
$M_{12}^{\rm FULL} = M_{12}^{\rm SM} + M_{12}^{\rm SUSY}$ whereas
$\Gamma_{12}^{\rm FULL} \approx \Gamma_{12}^{\rm SM}$.  In this case,
the dilepton asymmetry could be approximated as
\begin{equation}
A_{ll} = {\rm Im} \left( { \Gamma_{12}^{\rm SM} \over
M_{12}^{\rm SM} + M_{12}^{\rm SUSY} } \right).
\end{equation}
In the SM, $\Gamma_{12}$ is given by \cite{Buras:1997fb}
\begin{eqnarray}
  \Gamma_{12}^{\rm SM} &=&
  (-1) \:
  \frac{G_F^2\:m_b^2\:M_{B_d}\:B_{B_d}\:f_{B_d}^2}{8\pi} \:
  \left[
    v_t^2 + \frac{8}{3}\:v_c\:v_t \left(
      z_c + \frac{1}{4} z_c^2 - \frac{1}{2} z_c^3
    \right) +
  \right. \nonumber\\
  & &
  \left.
    v_c^2
    \left\{
      \sqrt{1 - 4 z_c} \left( 1 - \frac{2}{3} z_c \right) +
      \frac{8}{3} z_c + \frac{2}{3} z_c^2 - \frac{4}{3} z_c^3 - 1
    \right\}
  \right] ,
\end{eqnarray}
where \(v_i \equiv V_{ib}\:V_{id}^* \) and \( z_c \equiv m_c^2 / m_b^2 \).
The minus sign comes from the convention of $CP$ operation on neutral $B$ 
mesons: $ CP \: | B_d^0 \rangle = - \: | \overline{B_d^0} \rangle $.
In Fig.~\ref{fig2},
we show the dilepton asymmetry $A_{ll}$ as functions of $\gamma$, 
after scanning over the allowed region of complex parameters, 
$( \delta_{13}^d )_{LL}$ and $( \delta_{13}^d )_{LR}$. 
The SM case 
is represented by the filled black rectangle. The region leading to too large
$B\rightarrow X_d \gamma$ is denoted by the light gray region,
and the dark gray region is with
$B(B\rightarrow X_d \gamma) < 1 \times 10^{-5}$. Note that the KM angle 
$\gamma$ cannot be arbitrary, but should be somewhere between 
$\sim 20^{\circ}$ and $80^\circ$ because of the $B\rightarrow X_d \gamma$ 
constraint. The resulting $A_{ll}$ is essentially the same as the SM 
predictions. Thus, one cannot expect a large deviation in $A_{ll}$ from 
the SM prediction in SUSY models with $S_3^3$ flavor symmetry group.

Now let us consider the branching ratio for a radiative decay 
$B \rightarrow X_d \gamma$ and direct CP violation therein. The relevant 
operators for this decay are the current-current $b\rightarrow d u \bar{u}$ 
and (chromo)magnetic dipole operators \cite{ali}:
\begin{eqnarray}
{\cal H}_{\rm eff} ( b \rightarrow d \gamma (+g)) 
& = & - {4 G_F \over \sqrt{2}}
V_{td}^* V_{tb}~\sum_{i=1,2,7,8}~C_i (\mu_b) O_{ic} (\mu_b)
\nonumber  \\
& + &   {4 G_F \over \sqrt{2}}
V_{ud}^* V_{ub}~\sum_{i=1,2}~C_i (\mu_b) ~
\left[ O_{iu} (\mu_b) - O_{ic} (\mu_b)  \right] 
\end{eqnarray}
with
\begin{eqnarray}
  O_{1c} = \overline{d}_L \gamma^{\mu} c_L~\overline{c}_L \gamma_\mu b_L ,
&& 
  O_{1u} = \overline{d}_L \gamma^{\mu} u_L~\overline{u}_L \gamma_\mu b_L ,
\nonumber 
\\
  O_{2c} = \overline{d}_L \gamma^{\mu} c_L~\overline{c}_L \gamma_\mu b_L ,
&& 
  O_{2u} = \overline{d}_L \gamma^{\mu} u_L~\overline{u}_L \gamma_\mu b_L ,
\nonumber
\\
O_{7\gamma} =  {e \over 16 \pi^2} m_b ~
\overline{d}_L \sigma^{\mu\nu} F_{\mu\nu}  b_R ,
&& 
%\nonumber
%\\
O_{8g}  =  {g_s \over 16 \pi^2} m_b ~
\overline{d}_L \sigma^{\mu\nu} t^a G_{\mu\nu}^a  b_R .
\end{eqnarray}
Here the renormalization scale $\mu_b$ is of the order of $m_b$, and 
we have used the unitarity of the CKM matrix elements 
\[
V_{cd}^* V_{cb} = - ( V_{ud}^* V_{ub} + V_{td}^* V_{tb}),
\]
which should be valid even in the presence of SUSY flavor violations. 

In the SM, all the three up-type quarks are relevant to this decay, since
all the relevant CKM factors are of the same order of magnitude.
The strong phases are provided by the imaginary parts of one loop diagrams
at the $O( \alpha_s )$ order by the usual unitarity argument. The resulting
% branching ratio for this decay is $6.0 \times 10^{-6} - 2.6 \times 10^{-5}$ 
branching ratio for this decay is $7.5 \times 10^{-6} - 1.3 \times 10^{-5}$
% branching ratio for this decay is $8.9 \times 10^{-6} - 1.1 \times 10^{-5}$
in the SM depending on the CKM elements, and the direct CP asymmetry is 
% about $-7 \% \sim -35 \%$ in the SM \cite{ali,keum}. 
about $-18\% \sim -8\%$ in the SM \cite{ali,keum}. 
% about $-15\% \sim -10\%$ in the SM \cite{ali,keum}. 

The CP averaged branching ratio for $B\rightarrow X_d \gamma$ in the leading
log approximation is given by  
\begin{equation}
{B ( B \rightarrow X_d \gamma ) \over B( B \rightarrow X_c e \nu ) } 
= \left| {V_{td}^* V_{tb} \over V_{cb}} \right|^2 ~{6 \alpha \over \pi f(z) }~
| C_7 ( m_b ) |^2. 
\end{equation} 
where $f(z) = 1 - 8 z + 8 z^3 - z^4 - 12 z^2 \ln z$ is the phase space factor
for the $b\rightarrow c$ semileptonic decays and $\alpha^{-1} = 137.036$.
Neglecting the RG running between heavy SUSY particles and top quark mass 
scale, we get the following relations :
\begin{eqnarray}
C_7 ( m_b ) & \approx & - 0.31 + 0.67 ~C_7^{\rm new} ( m_W ) + 
0.09 ~C_8^{\rm new} ( m_W ), 
\nonumber   \\
C_8 ( m_b ) & \approx & - 0.15 + 0.70~ C_8^{\rm new} ( m_W ). 
\end{eqnarray}
The new physics contributions to $C_2$ are negligible so that we use 
$C_2 ( m_b )= C_2^{\rm SM} ( m_b ) \approx 1.11$. 
 
The direct CP asymmetry can be written as
\begin{eqnarray}
A_{\rm CP}^{b\rightarrow d\gamma} ({\rm in} \% ) & = & 
{1\over |C_7|^2}~\left[ 10.57 ~{\rm Im} \left( C_2 C_7^* \right) 
- 9.40 ~{\rm Im} \left( ( 1 + \epsilon_d ) C_2 C_7^* \right)   \right.
\nonumber   \\ 
& - & \left. 
9.51 ~{\rm Im} \left( C_8 C_7^* \right) + 0.12 ~{\rm Im} \left(
( 1 + \epsilon_d ) C_2 C_8^* \right)
\right]  
\end{eqnarray}
where 
\[
\epsilon_d \equiv {V_{ud}^* V_{ub} \over V_{td}^* V_{tb}}  
\approx { ( \rho - i \eta ) \over  ( 1 - \rho + i \eta )}.
\] 
A remark is in order for the above CP asymmetry in $B\rightarrow X_d \gamma$.
Unlike the $B\rightarrow X_s \gamma$ case for which the $|C_{7\gamma}|$ is 
constrained by the observed  $B\rightarrow X_s \gamma$ branching ratio, 
the  $B\rightarrow X_d \gamma$ decay has not been observed yet, and its 
branching ratio can be vanishingly small in the presence of new physics.
In that case, $|C_{7\gamma}| \approx 0$ so that the denominator of 
$A_{\rm CP}^{b\rightarrow d\gamma}$ becomes zero and the CP asymmetry blows 
up. This could be partly cured by replacing the denominator $|C_{7\gamma}|^2$
by $K_{\rm NLO} (\delta)$ defined in Ref.~\cite{kn}:
\begin{eqnarray}
K_{\rm NLO} (\delta) ({\rm in} \%) 
& = & 0.11 | C_2 |^2 + 68.13 | C_7 |^2 + 0.53 | C_8 |^2 
%\nonumber  \\
- 16.55 {\rm Re}( C_2 C_7^* ) 
\nonumber  \\
& - & 0.01 {\rm Re}( C_2 C_8^* ) + 8.85  {\rm Re}( C_7 C_8^* ) 
+ 3.86  {\rm Re}( C_7^{(1)} C_7^* )
\end{eqnarray} 
for the photon energy cutoff factor $\delta = 0.3$. Here $C_7^{(1)}$ is the 
next-to-leading order contribution to $C_{7\gamma} (m_b)$:
\begin{equation}
C_{7\gamma}^{(1)} \approx 0.48 - 2.29 C_7^{\rm new} ( m_W ) 
- 0.12 C_8^{\rm new} ( m_W ).
\end{equation}
%Also we adopt the upper limit on $B( B \rightarrow X_s g) < 6.8 \%$ as 
%the upper limit on $B( B \rightarrow X_d g)$ in order to put an upper limit 
%on  $|C_8 ( m_b )|^2$:
%\begin{equation}
%B( B \rightarrow X_d g) = \lambda^2 \left[ ( 1 - \rho )^2 + \eta^2 \right] 
%~| C_8 ( m_b ) |^2~B ( B \rightarrow X_c e \nu ). 
%
%\end{equation} 

\begin{figure}
% \parbox{3in}{\epsfxsize=7.5cm \epsfbox{.eps}\\
% \centerline{\footnotesize(a) Branching ratio}}%
% \hspace{.2in}%
% \parbox{3in}{\epsfxsize=7.5cm \epsfbox{.eps}\\
% \centerline{\footnotesize(b) Direct CP asymmetry}}\\%
% \vspace{.2in}%
%\centerline{\epsfxsize=10cm \epsfbox{}}
  \subfigure[$B(B \rightarrow X_d \gamma)$]{%
    \includegraphics[width=8cm]{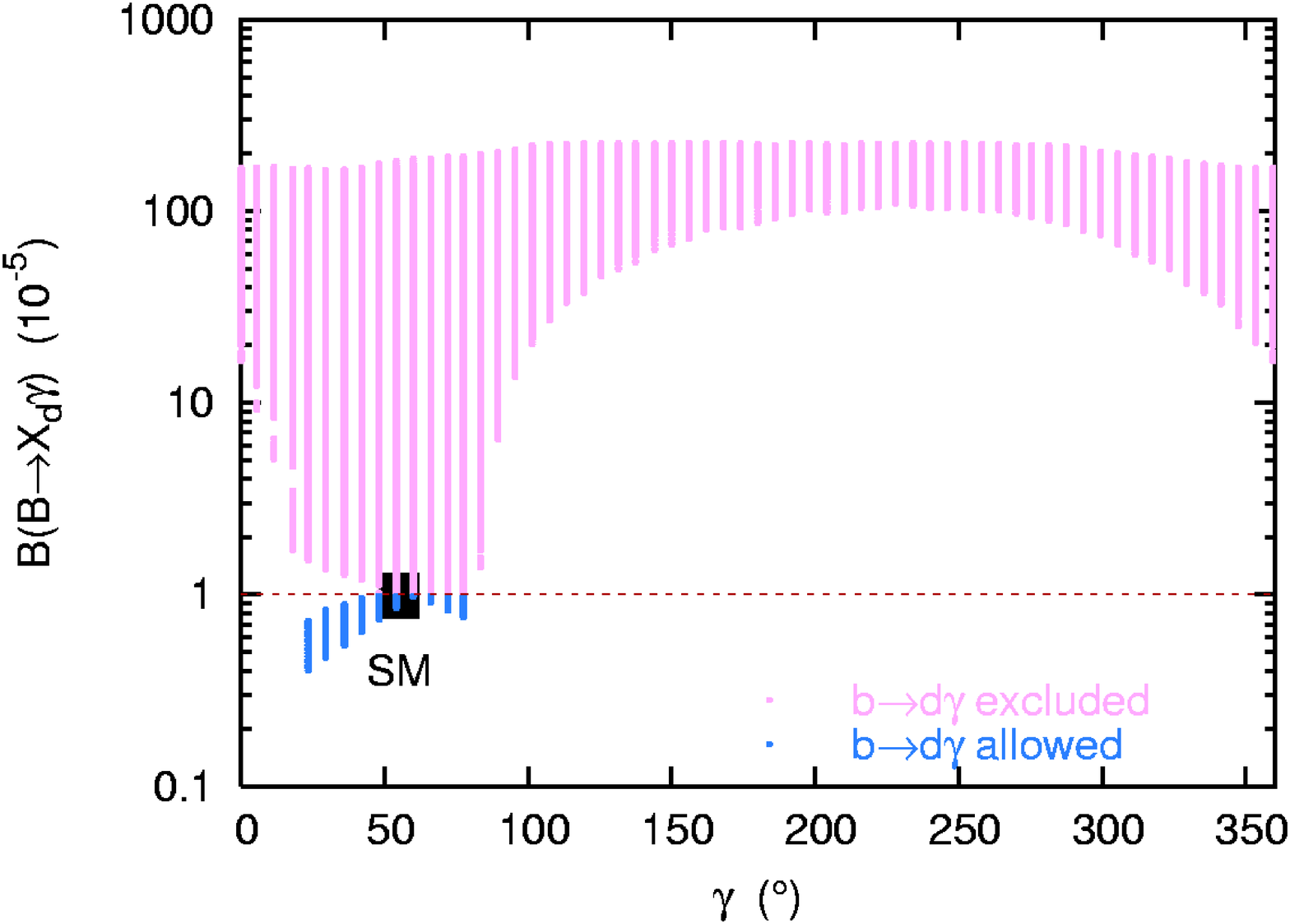}}
  \subfigure[$A_{\rm CP}^{b\rightarrow d\gamma}$]{%
    \includegraphics[width=8cm]{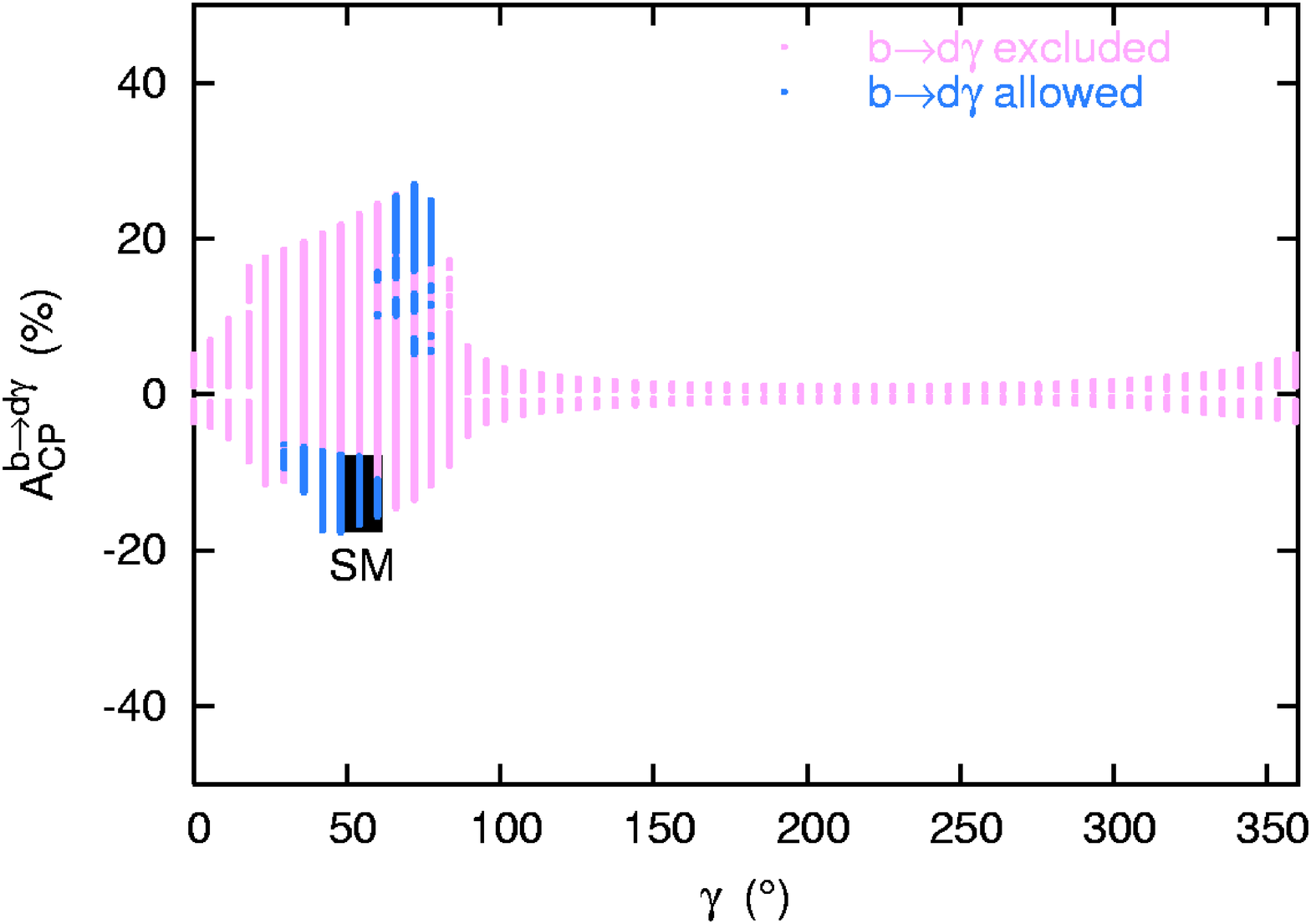}}
\caption{(a) Branching ratio and (b) direct CP asymmetry of 
$B \rightarrow X_d \gamma$ in our model as functions of the KM angle $\gamma$. 
The SUSY contributions are added to the SM contributions, and the induced 
$LR$ mixing is ignored. 
  The black rectangle around $\gamma \simeq 55^\circ$ is
  the SM prediction.
Those parameters which lead to
% $B ( B \rightarrow X_d \, g ) > 6.8 \%$ 
$B ( B \rightarrow X_d \gamma ) > 1 \times 10^{-4}$ 
are represented by
light gray (magenta), 
% $B ( B \rightarrow X_d \, g) < 6.8 \%$ by
% dark gray (green),
and $B ( B \rightarrow X_d \gamma ) < 1 \times 10^{-4}$ by
dark gray (blue).
}
\label{fig3}
\end{figure}

\begin{figure}
% \parbox{3in}{\epsfxsize=7.5cm \epsfbox{.eps}\\
% \centerline{\footnotesize(a) Branching ratio}}%
% \hspace{.2in}%
% \parbox{3in}{\epsfxsize=7.5cm \epsfbox{.eps}\\
% \centerline{\footnotesize(b) Direct CP asymmetry}}\\%
% \vspace{.2in}%
%\centerline{\epsfxsize=10cm \epsfbox{}}
  \subfigure[$B(B \rightarrow X_d \gamma)$]{%
    \includegraphics[width=8cm]{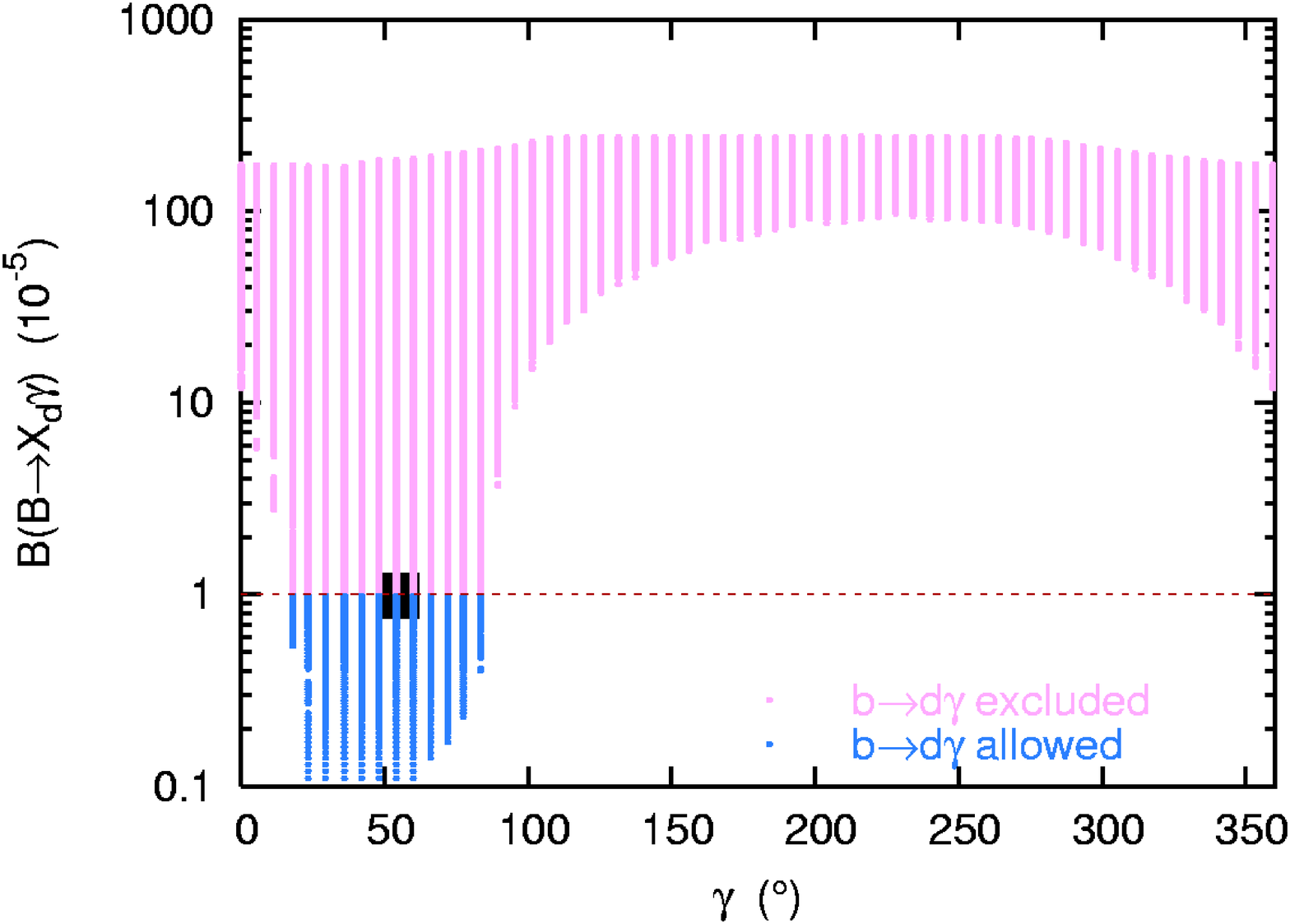}}
  \subfigure[$A_{\rm CP}^{b\rightarrow d\gamma}$]{%
    \includegraphics[width=8cm]{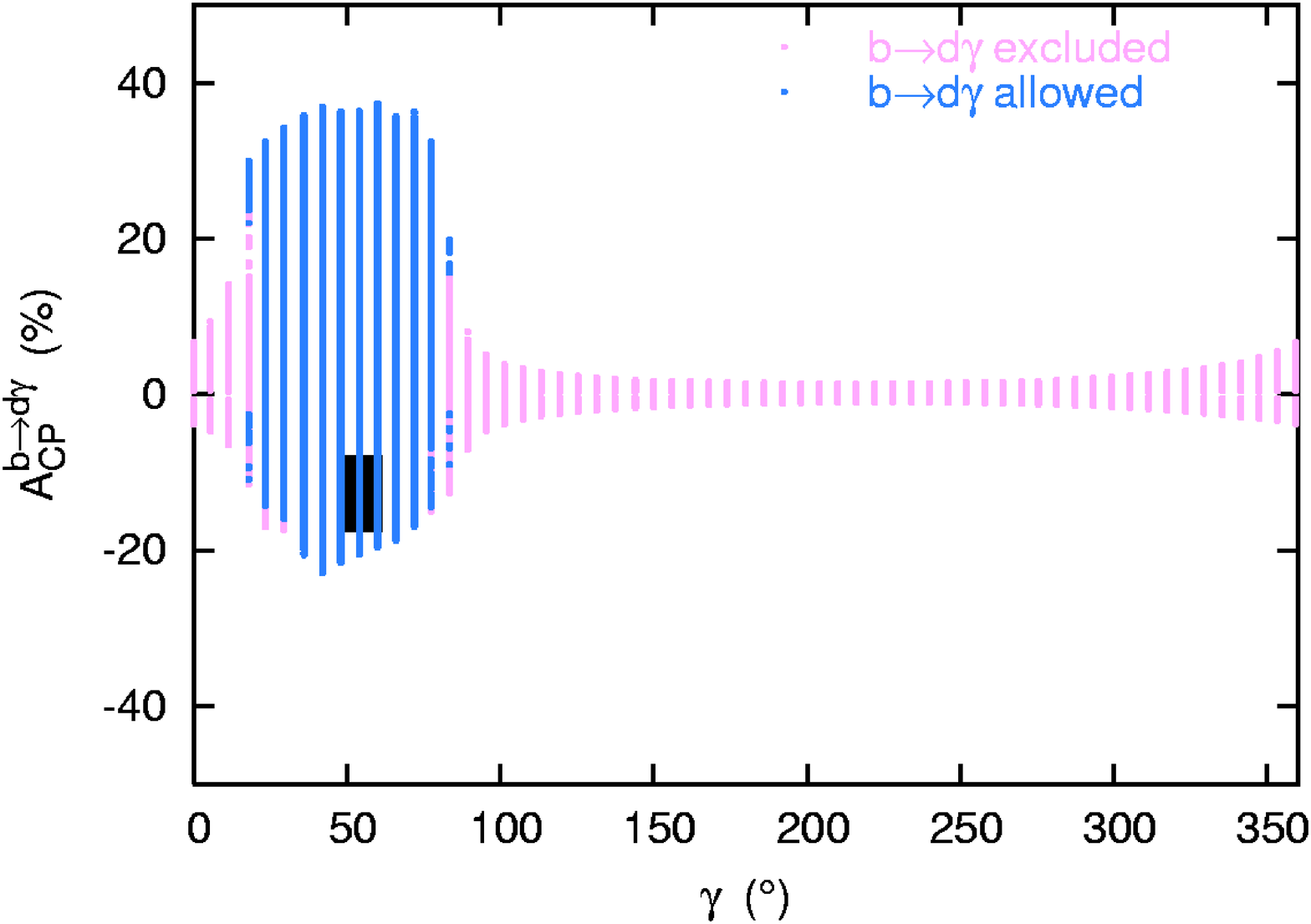}}
\caption{(a) Branching ratio and (b) direct CP asymmetry of 
$B \rightarrow X_d \gamma$ in our model as functions of the KM angle $\gamma$. 
The SUSY contributions are added to the SM contributions, and the induced
$LR$ mixing is included by setting $\mu \tan\beta = + 5$ TeV. 
  The black rectangle around $\gamma \simeq 55^\circ$ is
  the SM prediction.
Those parameters which lead to
% $B ( B \rightarrow X_d \, g ) > 6.8 \%$ 
$B ( B \rightarrow X_d \gamma ) > 1 \times 10^{-5}$ 
are represented by
light gray (magenta), 
% $B ( B \rightarrow X_d \, g) < 6.8 \%$ by
% dark gray (green),
and $B ( B \rightarrow X_d \gamma ) < 1 \times 10^{-5}$ by
dark gray (blue).
}
\label{fig4}
\end{figure}

In our model, the Wilson coefficients $C_{7\gamma}$ and $C_{8 g}$ are 
modified in the double mass insertion approximation as follows 
\cite{ko2,gabriel,calculations} :
\begin{eqnarray}
  C_{7 \gamma}^{\rm SUSY} ( m_W ) & = &  
{8 \pi Q_b \alpha_s \over 3 \sqrt{2} G_F \tilde{m}^2 V_{td}^* V_{tb}}
\left[ ( \delta_{13}^d )_{LL}  M_4 (x)  \right.
\nonumber  \\
& & \left. 
- ( \delta_{13}^d )_{LR} \left( {\tilde{m} \sqrt{x} \over m_b} \right) 
4 B_1 (x) 
- ( \delta_{13}^d )_{LR}^{\rm ind}
\left( {\tilde{m} \sqrt{x} \over m_b} \right) M_2 (x) \right],
\\
  C_{8 g}^{\rm SUSY} ( m_W ) & = &  {2 \pi \alpha_s \over 
\sqrt{2} G_F \tilde{m}^2 V_{td}^* V_{tb}}
\left[ ( \delta_{13}^d )_{LL} \left( {3\over 2} M_3 (x) - {1\over 6} M_4 (x)
\right) \right.
\nonumber  \\
& & 
+ ( \delta_{13}^d )_{LR} \left( {\tilde{m} \sqrt{x} \over m_b} \right) 
~{1\over 6}~\left( 4 B_1 (x) - 9 x^{-1} B_2 (x) \right)  
\nonumber  \\
& & \left. 
- ( \delta_{13}^d )_{LR}^{\rm ind}
\left( {\tilde{m} \sqrt{x} \over m_b} \right)
\left( {3\over 2} M_1 (x) - {1\over 6} M_2 (x) \right) \right].
\end{eqnarray}
Note that both $(\delta_{13}^d )_{LR}$ and $(\delta_{13}^d )_{LR}^{\rm ind}$
are enhanced by $m_{\tilde{g}} / m_b$ due to the chirality flip from the 
internal gluino propagator in the loop. Explicit expressions for the loop 
functions $B_i$'s and $M_i$'s can be found in Ref.~\cite{ko2}. 
In Figs.~\ref{fig3} (a) and (b), 
we show the branching ratio and the direct asymmetry 
of $B \rightarrow X_d \gamma$ as functions of $\gamma$ when the induced $LR$
mixing can be neglected. 
In Fig.~\ref{fig4} (a) and (b), we show the same plots when 
the induced $LR$ mixing is important by fixing $\mu \tan\beta = + 5$ TeV. 
First of all, the branching ratio for $B\rightarrow X_d \gamma$ comes out 
too large for $( \delta_{13}^d )_{LR} \sim \text{(a~ few)} \times 10^{-2}$,
compared to the well measured $B\rightarrow X_s \gamma$ : 
$B ( B \rightarrow X_s \gamma ) = (3.21 \pm 0.43_{stat} \pm 
0.27^{+0.18}_{(sys)-0.10 (th)} ) \times 10^{-4}$. Although there is no 
reported upper limit on $B\rightarrow X_d \gamma$ at the moment, it would be
reasonable to assume $B ( B\rightarrow X_d \gamma ) \lesssim 
1 \times 10^{-5}$, as discussed in the introduction. 
Imposing this condition, substantial 
parameter space in the $( LL, LR)$ plane is excluded, and the KM angle 
$\gamma$ cannot be too much different from $\gamma \sim \gamma_{SM} \sim 
55^{\circ}$. This would imply that the kaon physics in this model would not
be too different from the SM case. Still, there is some room that 
$ B( B\rightarrow X_d \gamma ) $ can differ from the SM case. 
Especially, the direct CP asymmetry in $B\rightarrow X_d \gamma$ can be 
significantly different from the SM prediction. % by a significant amount.
Even its sign can change from the SM case. Therefore the branching ratio of
$B\rightarrow X_d \gamma$ and direct CP asymmetry therein could be a 
sensitive probe for gluino mediated $b\rightarrow d$ transitions in 
$B^0 - \overline{B^0}$ mixing and $b\rightarrow d\gamma$, and will provide 
an important information for the flavor structure of down squark mass 
matrices. 

In conclusion, we presented a simple SUSY model where SUSY flavor problem is 
solved by $S_3^3$ flavor symmetry group, in which the mass insertion parameters
have typical sizes shown in Table~1. In the kaon sector $(\delta_{12}^d)_{LL}$ 
insertion is dominant over other parameters, and it is possible to have large
SUSY contributions to $\epsilon_K$ and $\epsilon^{'}/\epsilon_K$ as discussed
in Refs.~\cite{ko,ko2} with unconstrained KM angle $\gamma$.  However, this
picture cannot be retained when CP violations in the $B$ meson sector is 
considered. Since $(\delta_{13}^d )_{LL}$ $\sim$ $(\delta_{13}^d )_{LR}$ 
$\sim$ $O( \lambda^3 )$, only $LR$ mixing can be important for $B^0 - 
\overline{B^0}$ mixing. However the $LR$ mixing is strongly constrained by
$B \rightarrow X_d \gamma$ which has not been observed yet. By imposing
a reasonable limit on $B ( B \rightarrow X_d \gamma) < 1 \times 10^{-5}$,
we find that $B^0 - \overline{B^0}$ mixing is dominated by the SM 
contributions, and not by SUSY contributions. The KM angle $\gamma$ must 
lie between $\sim 20^\circ$ and $\sim 80^\circ$. Still a small amount of 
$(\delta_{13}^d )_{LR}$ can give a significant contribution to 
$B\rightarrow X_d \gamma$ and direct CP asymmetry therein. One can still 
expect large deviations in these observables in SUSY models with flavor 
$S_3^3$ symmetry. Since the KM angle $\gamma$ is not free, the resulting 
predictions for $K \rightarrow \pi\nu\bar{\nu}$ and  
$K_L \rightarrow \pi^0 e^+ e^-$ will be constrained as well within the 
predictions given in Ref.~\cite{ko2}.
If the low energy SUSY is relevant to the nature, some mechanism is needed 
to solve the SUSY flavor/CP problems. If the approximate alignment based 
on $S_3^3$ is such a solution, one 
expects generically large modifications in $K^0 - \overline{K^0}$, 
$K\rightarrow \pi \nu \bar{\nu}$, $K_L \rightarrow \pi^0 e^+ e^-$ and 
$B\rightarrow X_d \gamma$, but not in $B^0 - \overline{B^0}$ mixing.
Detailed experimental search for these decays at $K$ and $B$ factories 
will shed light on the flavor physics in SUSY models.

% figures should be put into the text as floats.
% Use the graphics or graphicx packages (distributed with LaTeX2e)
% and the \includegraphics macro defined in those packages.
% See the LaTeX Graphics Companion by Michel Goosens, Sebastian Rahtz,
% and Frank Mittelbach for instance.
%
% Here is an example of the general form of a figure:
% Fill in the caption in the braces of the \caption{} command. Put the label
% that you will use with \ref{} command in the braces of the \label{} command.
% Use the figure* environment if the figure should span across the
% entire page. There is no need to do explicit centering.

% \begin{figure}
% \includegraphics{}%
% \caption{\label{}}
% \end{figure}

% Surround figure environment with turnpage environment for landscape
% figure
% \begin{turnpage}
% \begin{figure}
% \includegraphics{}%
% \caption{\label{}}
% \end{figure}
% \end{turnpage}

% Specify following sections are appendices. Use \appendix* if there
% only one appendix.
%\appendix
%\section{}

\begin{acknowledgments}
We are grateful to Seungwon Baek for useful discussions. 
A part of this work was done at Aspen Center for Physics in the fall of 2002. 
This work was supported by BK21 Haeksim program of Ministry of Education, 
SRC program of KOSEF through CHEP at Kyungpook National University, 
and DFG-KOSEF Collaboration program under the contract
20005-111-02-2 (KOSEF) and 446 KOR-113/137/0-1 (DFG) (JHP).
\end{acknowledgments}

% Create the reference section using BibTeX:
%\bibliography{basename of .bib file}

\end{document}